\documentclass[11pt,a4paper]{article}
\usepackage{jcappub}

\usepackage{bm}
\usepackage{epsfig}
\usepackage{citesort}
\usepackage{graphicx}
\usepackage{amsmath}
\usepackage{amssymb}
\usepackage{amsbsy}
\usepackage{color}
\usepackage{subfigure}
\usepackage{slashed}
\usepackage{afterpage}
\usepackage{psfrag}
\usepackage{axodraw}
\usepackage{graphicx}
\usepackage{color}
\usepackage{amsmath}
\usepackage{soul}

\newcommand{\be}{\begin{equation}}
\newcommand{\ee}{\end{equation}}
\newcommand{\bea}{\begin{eqnarray}}
\newcommand{\eea}{\end{eqnarray}}

\newcommand{\I}{{\rm i}}

\begin{document}
\subheader{\hfill MPP-2013-110}

\title{Solar constraints on hidden photons re-visited}

\author[a,b]{Javier~Redondo}
\author[a]{and Georg~Raffelt}

\affiliation[a]{Arnold Sommerfeld Center,
Ludwig-Maximilians-Universit\"at, Theresienstr.~37,\\ 80333
M\"unchen, Germany}

\affiliation[b]{Max-Planck-Institut f\"ur Physik
(Werner-Heisenberg-Institut), F\"ohringer Ring 6,\\ 80805 M\"unchen,
Germany} \emailAdd{redondo@mpp.mpg.de}

\abstract{ We re-examine solar emission of hidden photons $\gamma'$
(mass $m$) caused by kinetic $\gamma$--$\gamma'$ mixing. We
calculate the emission rate with thermal field theory methods and
with a kinetic equation that includes $\gamma$--$\gamma'$ ``flavor
oscillations'' and $\gamma$ absorption and emission by the thermal
medium. In the resonant case both methods yield identical emission
rates which, in the longitudinal channel, are enhanced by a factor
$\omega_{\rm P}^2/m^2$ (plasma frequency $\omega_{\rm P}$) in
agreement with An, Pospelov and Pradler (2013). The Sun must not
emit more energy in a ``dark channel'' than allowed by solar
neutrino measurements, i.e., not more than 10\% of its photon
luminosity. Together with the revised emission rate, this
conservative requirement implies $\chi<4\times 10^{-12} ({\rm
eV}/m)$ for the kinetic mixing parameter. This is the most
restrictive stellar limit below  $m \sim 3~{\rm eV}$,
whereas for larger masses the transverse channel dominates together
with limits from other stars. A recent analysis of XENON10 data
marginally improves the solar limit, leaving open the opportunity to
detect solar hidden photons with future large-scale dark matter
experiments. Detecting low-mass hidden photons with the ALPS-II
photon-regeneration experiment also remains possible. }

\maketitle

\section{Introduction}

At the low-energy frontier of particle
physics~\cite{Jaeckel:2010ni}, the existence of hidden photons (HPs)
is an intriguing hypothesis. These particles would have a small mass
$m$ and would be completely sterile except for kinetic mixing with
normal photons, described by a dimensionless parameter $\chi$. The
Lagrangian describing the coupled system is
\begin{equation}\label{eq:lagrangian}
{\cal L}=-{\frac{1}{4}} A_{\mu\nu}A^{\mu\nu}
-{\frac{1}{4}} B_{\mu\nu}B^{\mu\nu}
+{\frac{m^2 }{2}}B_{\mu}B^{\mu}
-{\frac{\chi}{2}}  A_{\mu\nu}B^{\mu\nu}\,,
\end{equation}
where $A$ and $B$ are the photon and HP fields.

Ongoing experimental searches look for photon appearance from
putative HP sources. This includes photon regeneration experiments,
beam dump experiments or searches for solar HPs. The usual
requirement that stars should not lose excessive amounts of energy
in dark channels also leads to restrictive limits~\cite{Okun:1982xi,
popov:1991, Popov1999, Redondo:2008aa, An:2013yfc}. HP production in
the early universe can be quite efficient despite the small
couplings, which makes them excellent candidates for dark
matter~\cite{Pospelov:2008jk, Redondo:2008ec, Nelson:2011sf,
Arias:2012az} and dark radiation~\cite{Jaeckel:2008fi} in different
parameter ranges. A recent review echoes the various
phenomenological aspects of HP models~\cite{Jaeckel:2013ija}.

Returning to stars, it is a peculiarity of HP emission that it
depends on density and temperature such that, for small masses,
the most restrictive limit arises from the Sun, whereas usually
globular cluster stars, white dwarfs, or neutron stars are more
constraining.

One way of looking at HP production in a hot medium is
\hbox{$\gamma$--$\gamma'$} oscillations in conjunction with $\gamma$
absorption and emission, similar to the production of sterile
neutrinos in the early universe by active-sterile oscillations. In
the stellar plasma, both transverse (T) and longitudinal (L)
electromagnetic excitations exist, and both mix with the
corresponding HP polarizations. In a nonrelativistic plasma,
L-plasmons have the approximate dispersion relation
$\omega=\omega_{\rm P}$ (plasma frequency) which to lowest order in
the small electron velocity does not depend on photon wavenumber
${k}$, whereas HPs obey $\omega^2=m^2+{k}^2$. Hence, for any value
of $\omega_{\rm P}$, i.e., at any location in the Sun, there is a
population of L-plasmons with the HP wavenumber $k^2= \omega^2-m^2$,
leading to maximal $\gamma$--$\gamma'$ mixing and thus to resonant
emission. This peculiarity of the L dispersion relation makes the L
channel important.

As a consequence of the unusual dispersion relation, it is also important
to include the non-trivial wave-function renormalization factor in
processes with external L-plasmons, corresponding to the correct
residue factor of the L-plasmon propagator. In the first calculation
of neutrino emission by plasmon decay~\cite{Adams:1963zzb},
$\gamma\to\nu\bar\nu$, a superfluous factor $K^2/\omega^2$ was
included for L-plasmon decay~\cite{Tsytovich:1963, Zaidi:1965}.
Here, $K=(\omega_{\rm P},{\bf k})$ is the four-momentum of the
decaying plasmon. Such incorrect residue factors can easily sneak
in, depending on the choice of gauge used in the calculation.
Plasmon decay was later extended to relativistic plasmas
\cite{Braaten:1990de, Itoh:1992}, and once more this factor has
crept in (see the remark after Eq.~A9 in
Ref.~\cite{Braaten:1993jw}). These errors had no practical
consequences because, after phase-space integration, the
modification was not dramatic and the L-plasmon channel was
subdominant anyway.

A similar error in the first calculation of HP emission by one of
us~\cite{Redondo:2008aa} once more leads to the same spurious
factor $K^2/\omega^2$, but this time with drastic
consequences~\cite{An:2013yfc}. This is because here $K^2=m^2$ for
the emitted HP so that the emission rate was underestimated by a
factor $m^2/\omega_{\rm P}^2$. With $\omega_{\rm P}\sim 0.3$~keV in
the Sun, the correction factor is huge for very low-mass HPs.

While it would be enough to re-scale the previous results with this
factor, another aspect of the enhanced emission rate is HP
production with keV energies that may be accessible in detectors,
but do not correspond to propagating L-plasmons in the Sun. The
oscillation mechanism is then not an adequate description. Instead,
one may think of the emission, for example, as bremsstrahlung with
the electromagnetic field mediating between the electrons and hidden
photons in the form of an off-shell propagator~\cite{An:2013yfc}.
This calculation includes, as a special case, the on-shell
production and must reproduce the results from the oscillation
approach.

This ``propagator approach'' simplifies even further if
one uses thermal field theory from the start. The calculation of the
emission rate then reduces to the almost trivial exercise of
reading the expression for the HP thermal self-energy from the
Lagrangian Eq.~(\ref{eq:lagrangian}). The role of the residue factor
also becomes immediately obvious.

For on-shell photon--HP conversion, the essential physics is that
thermal plasmons oscillate into HPs, collisions with the medium
absorb the photon component, destroy the coherence of the mixed
state, and in this way produce a population of sterile states that
escape. In the ``propagator treatment,'' oscillations never appear
explicitly and one may wonder about its range of validity.

The physics of flavor oscillations combined with collisions is
captured in a simple Boltzmann collision equation originally devised
for neutrinos~\cite{Sigl:1992fn}. To make connection between the
different approaches we also derive HP emission in this picture. In
the end we find identical results for the physical circumstances at
hand. The crucial condition under which both approaches reconcile is
the average flavor conversion between collisions being small. This
can happen in two radically different situations: if the mixing
angle is small or, even for a large mixing angle, the collisions
happen so fast that oscillations do not get far between collisions
(strong damping or quantum-Zeno regime). Both conditions are
satisfied if the active-sterile ``mixing energy'' is small compared
to the interaction rate of the active quanta. In our case, the
mixing energy in the L-channel is $\frac{1}{2}\chi m$. This quantity
is extremely small compared with the relevant L-plasmon absorption
rate: we are deeply in the ``allowed regime'' for the thermal field
theory treatment.

\section{Thermal field theory derivation}

\label{sec:thermalapproach}

To study the photon--HP system described by the Lagrangian
Eq.~(\ref{eq:lagrangian}), usually the non-orthogonal fields $A$
and $B$ are redefined by $B_\mu=S_\mu-\chi A_\mu$ to introduce the
sterile field $S_\mu$. The mixing is thus shifted from the kinetic
term to a matrix of mass-squares that mixes the flavors $S$ and $A$
and we find
\begin{equation}\label{eq:lagrangianS}
{\cal L}=-{\frac{1}{4}} A_{\mu\nu}A^{\mu\nu}
-{\frac{1}{4}} S_{\mu\nu}S^{\mu\nu}
+{\frac{m^2 }{2}}(S_\mu-\chi A_\mu)^2\,.
\end{equation}
We are interested in small $\chi$ and derive our formulas
perturbatively in $\chi$. $S$ quanta are produced at order $\chi^2$,
while distinguishing between $B$ and $S$ introduces corrections of
order $\chi^4$ so that in practice we may identify $S$ with $B$.

To obtain the HP emission rate we first observe that the emission
rate by a thermal medium for any particle is closely related to the
imaginary part of its self-energy $\Pi$ in the medium. In particular
\cite{Weldon:1983jn}
\begin{equation}\label{eq:imgamma}
{\rm Im}\,\Pi=-\omega\,\Gamma\,,
\end{equation}
where $\Gamma$ is the rate by which the particle distribution
approaches thermal equilibrium. We also consider the absorption rate
$\Gamma_{\rm abs}$ (inverse mean free path) and the spontaneous
production rate $\Gamma_{\rm prod}$. Detailed balancing reveals
$\Gamma_{\rm prod}=e^{-\omega/T}\,\Gamma_{\rm abs}$. Moreover, for
bosons $\Gamma=\Gamma_{\rm abs}-\Gamma_{\rm
prod}=(e^{\omega/T}-1)\,\Gamma_{\rm prod}$. The desired thermal
production rate is thus found to be
\begin{equation}\label{eq:gammaprodimpi}
\Gamma_{\rm prod}=-\frac{{\rm Im}\,\Pi}{\omega\,(e^{\omega/T}-1)}\,.
\end{equation}

\begin{figure}[b]
\centering
\includegraphics[width=0.5\columnwidth]{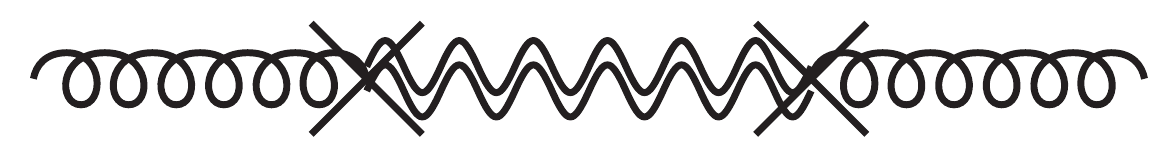}
\caption{Self-energy for hidden photons (curly lines) in a thermal medium
by mixing with thermal photons of the medium (double line).}\label{fig:graph}
\end{figure}

We thus seek the imaginary part of the HP self-energy in the medium.
At order $\chi^2$ it is given by the graph of Fig.~\ref{fig:graph}
after truncating the external lines. The crosses symbolize the
mixing vertex, the internal double line is the thermal photon
propagator. Without further ado we find the HP self-energy for a
given polarization state to~be
\begin{equation}
\label{eq:Sselfenergy}
\Pi_S(K)=m^2+\chi m^2\,\frac{1}{K^2-\Pi_A(K)}\,\chi m^2\,,
\end{equation}
where we use Lorentz gauge and $K=(\omega,k)$ is the four-momentum
of the external HP. It fulfills $K^2={\rm Re}\,\Pi_S=m^2$ up to
corrections of order $\chi^2$, which disappear in the $m\to 0$ limit
as seen in Eq.~(\ref{eq:Sselfenergy}).  The real part of the photon
self-energy can be taken from the literature. For the imaginary part
we can use Eq.~(\ref{eq:imgamma}) with $\Gamma_A$ derived from
electron-nucleus bremsstrahlung emission or other processes. The HP
production rate thus follows directly from that for ordinary
photons.

In a nonrelativistic plasma, one finds to lowest order in the small
electron velocity
\begin{equation}\label{eq:polfunction}
{\rm Re}\,\Pi_{\rm T}=\omega_{\rm P}^2
\quad\hbox{and}\quad
{\rm Re}\,\Pi_{\rm L}=\frac{K^2}{\omega^2}\,\omega_{\rm P}^2\,.
\end{equation}
The plasma frequency is $\omega_{\rm P}^2 =4\pi\alpha\,n_e/m_e$ with
$n_e$ the electron density. Corrections from the electron velocity
are of order $T/m_e$. With $T<1.3$~keV in the Sun, this is
negligible. The corresponding dispersion relations are
$\omega^2-k^2=\omega_{\rm P}^2$ for T-plasmons which therefore
behave exactly like particles with mass $\omega_{\rm P}$. L-plasmons
obey $\omega^2=\omega_{\rm P}^2$, independently of wavenumber.

For the L-channel we note that in the denominator $K^2=Z_{\rm
L}^{-1}\omega^2$ where we define $Z_{\rm L}=\omega^2/K^2$ so that
overall
\begin{equation}
{\rm Im}\,\Pi^S_{\rm L}=\chi^2 m^4\,{\rm Im}\,
\frac{Z_{\rm L}}{\omega^2-\omega_{\rm P}^2-\I Z_{\rm L}{\rm Im}\,\Pi_{\rm L}^A}\,.
\end{equation}
We recognize $Z_{\rm L}$ as the residue factor chosen to obtain the
canonical pole structure. With $K^2=m^2$ we have $Z_{\rm
L}=\omega^2/m^2$, explaining the enhanced L-emission noted in
Ref.~\cite{An:2013yfc}.

The residue factor can be interpreted as renormalizing ``charges''
to which the propagator is attached, here $\chi\to\sqrt{Z_{\rm
L}}\,\chi$. This is analogous to $\sqrt{Z_{\rm L}}$ renormalizing
the electron charge in plasmon decay. In the spirit of 
Eq.~(\ref{eq:imgamma}) we must interpret $-Z_{\rm L}{\rm
Im}\Pi_{\rm L}/\omega$ as the damping rate of L
quanta, which we denote as $\Gamma_{\rm L}$. 
On the pole of the propagator this is the \emph{physical}
damping rate of on-shell L-plasmons. It therefore
involves $Z_{\rm L}$, now renormalizing the electron charge, for
example in bremsstrahlung emission.\footnote{In the original HP
emission calculation~\cite{Redondo:2008aa}, $Z_{\rm L}$ was
correctly included in the damping-rate calculation, but it was
overlooked that $\chi$ should have been renormalized as well.}

Collecting all factors, we find the HP production rate in the
L-channel to be~\cite{An:2013yfc}
\begin{equation}\label{eq:Lproduction}
\Gamma^{\rm prod}_S=\frac{\chi^2 m^2}{e^{\omega/T}-1}\,\,
\frac{\omega^2\Gamma_{\rm L}}{(\omega^2-\omega_{\rm P}^2)^2+(\omega \Gamma_{\rm L})^2}
\,.
\end{equation}
Plasmons are weakly damped ($\Gamma_{\rm L}\ll\omega_{\rm P}$) so
that for $\omega \sim \omega_{\rm P}$ this is a narrow resonance
which, for the purpose of phase-space integration, is a delta
function
\begin{eqnarray}\label{eq:deltaapprox}
\Gamma^{\rm prod}_S
&\sim&\frac{\chi^2 m^2}{e^{\omega/T}-1}\,\frac{\pi}{2}\,
\delta(\omega-\omega_{\rm P})\,.
\end{eqnarray}
Remarkably, for resonant longitudinal HP emission, we do not need
any production rate. However, Eq.~(\ref{eq:Lproduction}) is valid
for any $\omega$ and thus can be used to compute the emission rate
for $\omega$ far away from the resonance. In this case one needs
$\Gamma_{\rm L}(\omega)=\Gamma_{\rm L}(K)|_{K^2=m^2}$ explicitly.

\section{Kinetic Approach}

Deriving the HP emission-rate in the above form is quick and elegant, but
at first glance its equivalence to the flavor oscillation approach
may not be entirely obvious. Therefore, we derive the emission rate
based on a kinetic equation usually employed in neutrino physics.

We begin with the equation of motion for the mixed fields, which is
in Fourier space~\cite{Redondo:2008aa}
\begin{equation}\label{eq:kleingordon1}
\left[\omega^2-k^2-
\begin{pmatrix}\pi(\omega,k)&-\chi m^2\\ -\chi m^2&m^2\end{pmatrix}
\right]
\begin{pmatrix}A\\ S\end{pmatrix}
=0\,,
\end{equation}
where $A$ and $S$ are the field amplitudes. We use the notation
$\pi(K)={\rm Re}\,\Pi(K)$ for the real part of the polarization
function which gives us the dispersion relation by virtue of
$\omega^2-k^2=\pi(\omega,k)$. These equations apply separately to
the L and T polarizations.

Oscillating particles are not in propagation eigenstates and thus do
not have simultaneously fixed energy and momentum. We here consider
evolution in time and thus use quanta of fixed common momentum $k$.
For simplicity we assume relativistic states so that $\omega\approx
k$.

For the T case we expand $\omega^2-k^2=(\omega-k)(\omega+k)\approx
(\omega-k)2k$. After linearizing the equation, we reverse the time
Fourier transform, i.e., $\omega\to \I\partial_{t}$ and find the
usual ``Schr\"odinger equation'' for flavor oscillations
\begin{equation}\label{eq:kleingordon2}
\I\partial_{t}
\begin{pmatrix}A(t)\\ S(t)\end{pmatrix}
=\begin{pmatrix}\omega_{A}&\mu\\ \mu&\omega_{S}\end{pmatrix}
\begin{pmatrix}A(t)\\ S(t)\end{pmatrix}\,.
\end{equation}
Here $\omega_{A,S}$ are the energies of $A$ and $S$ quanta following
from the dispersion relation for assumed momentum $k\approx\omega$
and $\mu=-\chi m^2/2k$ is a ``mixing energy.''

For L-modes the situation is more complicated because the
polarization function given in Eq.~(\ref{eq:polfunction}) is not
simply an effective mass but rather $\pi_{\rm
L}(\omega,k)=\omega_{\rm P}^2/Z_{\rm L}$ with $Z_{\rm
L}=\omega^2/K^2$. The Klein-Gordon equation is explicitly
\begin{equation}\label{eq:kleingordon3}
\begin{pmatrix}(\omega^2-\omega_{\rm P}^2)\,Z_{\rm L}^{-1}(K)&-\chi m^2\\
-\chi m^2&\omega^2-\omega_S^2\end{pmatrix}
\begin{pmatrix}A\\ S\end{pmatrix}
=0\,.
\end{equation}
It is brought to canonical form by $A\to A/\sqrt{Z_{\rm L}}$ and
$\chi\to \chi\,\sqrt{Z_{\rm L}}$ so that
\begin{equation}\label{eq:kleingordon4}
\begin{pmatrix}\omega^2-\omega_{\rm P}^2&-\chi m\omega\\
-\chi m\omega&\omega^2-\omega_S^2\end{pmatrix}
\begin{pmatrix}A\\ S\end{pmatrix}
=0\,.
\end{equation}
Linearizing it we recover an oscillation equation of the form
Eq.~(\ref{eq:kleingordon2}) with $\omega_{A}=\omega_{\rm P}$,
$\omega_{S}=(m^2+k^2)^{1/2}$, and $\mu=-\chi m/2$. The mixing energy
is enhanced by $\sqrt{Z_{\rm L}}$ in full analogy to the
propagator approach.\footnote{In the original
treatment~\cite{Redondo:2008aa}, the on-shell value
$\pi_{L}=\omega_{\rm P}^2-k^2$ was inserted in
Eq.~(\ref{eq:kleingordon1}), representing the correct dispersion
relation. However, in this case the external ``charge'' $\chi$
should have been renormalized with $\sqrt{Z_{\rm L}}$. This error
was at least partly caused by the presentation of $\pi_{\rm
L}(\omega,k)$ in Refs.~\cite{Haft:1993jt, Raffelt:1996} which is
only correct on-shell, i.e., for $\omega$ and $k$ connected by the
dispersion relation. The renormalization factors were separately
provided.}

The key ingredient for HP production is that flavor oscillations
described by this equation are interrupted by collisions and we must
understand the evolution of the ensemble, not of individual
particles. Therefore, the adequate description is in terms of
density matrices. For a fixed momentum ${\bf k}$, the free evolution
is described by a Hamiltonian for two coupled harmonic oscillators,
$H=\sum_{i,j=A,S} a_i^\dagger \Omega_{ij} a_j$. Here $a_i^\dagger$
and $a_i$ are the creation and annihilation operators, respectively,
of quanta with flavor $i$ and
\begin{equation}
\Omega=
\begin{pmatrix}
\omega_A&\mu\\ \mu&\omega_S
\end{pmatrix}
=\frac{\omega_A+\omega_S}{2}+\begin{pmatrix}
\frac{1}{2}\Delta\omega&\mu\\ \mu&-\frac{1}{2}\Delta\omega
\end{pmatrix}
\,,
\end{equation}
where $\Delta\omega=\omega_A-\omega_S$. The ``mixing energy'' $\mu$
is assumed to be small compared to the diagonal elements, causing
only a small overall energy modification.

In a kinetic approach, the evolution of the fields $A$ and $S$ is
described by the expectation values of field bilinears. In the
simplest case, the matrix of occupation numbers $ \rho_{ij}=\langle
a^\dagger_j a_i\rangle$ captures all relevant information. The field
$A$ is assumed to interact with the external medium by an
interaction linear in $A$, i.e., the medium can absorb or emit
$A$-quanta, but will not scatter them between different momenta. In
this case the evolution of different momentum modes is not coupled
and the equation of motion for a single momentum mode
is~\cite{Sigl:1992fn}
\begin{equation}
\dot\rho=-\I\,[\Omega,\rho]+{\textstyle\frac{1}{2}}
\{G_{\rm prod},1\pm\rho\}-{\textstyle\frac{1}{2}}\{G_{\rm abs},\rho\}\,,
\end{equation}
where $\{\cdot,\cdot\}$ is an anticommutator. The positive sign
applies to bosons (Bose stimulation), whereas the negative sign
applies to fermions (Pauli blocking). We have defined the matrices
\begin{equation}
G_{\rm prod}=
\begin{pmatrix}
\Gamma_{\rm prod}&0\\0&0
\end{pmatrix}
\quad\hbox{and}\quad
G_{\rm abs}=
\begin{pmatrix}
\Gamma_{\rm abs}&0\\0&0
\end{pmatrix}\,,
\end{equation}
where $\Gamma_{\rm prod}$ and $\Gamma_{\rm abs}$ are the production
and absorption rates of $A$-quanta with momentum ${\bf k}$ by the
medium. In thermal equilibrium they obey $\Gamma_{\rm
prod}=e^{-\omega_A/T} \Gamma_{\rm abs}$. The commutator part
describes flavor oscillations and is equivalent to the earlier
Schr\"odinger equation.

In thermal equilibrium and ignoring flavor oscillations, the
$S$-particles will not be excited at all, whereas the $A$-particles
are thermally occupied according to
$f_{T}=(e^{\omega_A/T}\pm1)^{-1}$, the negative sign applying to
bosons, the positive one to fermions. We then describe a
non-equilibrium situation by its deviation from equilibrium in the
form
\begin{equation}
\rho=\rho_{T}+\delta\rho=
\begin{pmatrix}f_{T}&0\\0&0\end{pmatrix}+\delta\rho\,.
\end{equation}
The collision term vanishes for $\rho_{T}$ and we are left with
\begin{equation}
\dot\rho=-\I\,[\Omega,\rho]-
{\textstyle\frac{1}{2}} \{G,\delta\rho\}\,,
\end{equation}
where $G={\rm diag}(\Gamma,0)$. The damping rate is
$\Gamma=\Gamma_{\rm abs}\pm\Gamma_{\rm prod} =(1\pm
e^{-\omega_A/T})\Gamma_{\rm abs}$. The negative sign is for bosons,
i.e., the distribution function approaches equilibrium with the
difference between the ``spontaneous'' absorption and emission rates
as mentioned earlier, whereas for fermions (positive sign) it is the
sum.

We next write the $\rho$ matrix in terms of occupation number
components explicitly
\begin{equation}
\delta\rho=
\begin{pmatrix}
h_A&g\\g^*&h_S
\end{pmatrix}\,,
\end{equation}
leading to the equations of motion
\begin{eqnarray}
\dot h_A&=&-\Gamma h_A-2\mu\,{\rm Im}(g)
\\\label{eq:hBmotion}
\dot h_S&=&2\mu\,{\rm Im}(g)
\\ \label{eq:gmotion}
\dot g&=&-\left({\textstyle \frac{1}{2}}\,\Gamma
+\I\,\Delta\omega\right) g
+\I \mu \left(f_{T}+h_A-h_S\right)\,.
\end{eqnarray}
In the absence of damping ($\Gamma=0$), these equations describe
flavor oscillation, in the absence of mixing ($\mu=0$), the approach
of $A$ to thermal equilibrium.

When $\Gamma\gg\mu$ the damping rate is much larger than the
oscillation frequency when mixing is maximal for $\Delta\omega=0$.
In other words, on resonance we are in the strong damping regime
(quantum Zeno regime), where decoherence between the mixed species
is faster than oscillations. When $\Delta\omega$ is sufficiently
large, this is no longer the case, but then mixing becomes small.
Thus our solution will never stray far from thermal equilibrium,
i.e.\ $|h_A|\ll f_{T}$ and $|h_S|\ll f_{T}$.

In this limit, Eq.~(\ref{eq:gmotion}) is a closed equation of motion
of the form $\dot g=-\left({\textstyle \frac{1}{2}}\,\Gamma
+\I\,\Delta\omega\right) g +\I \mu f_{T}$. With the initial
condition $g(0)=0$ it has the solution
\begin{equation}\label{eq:gmotion3}
g(t)=\frac{1-e^{-(\I\,\Delta\omega+\Gamma /2)t}}{\Delta\omega-\I \Gamma/2}\,
\mu f_{T}\,.
\end{equation}
After an initial transient it approaches the steady-state solution
\begin{equation}\label{eq:gmotion4}
g_\infty=
\frac{\Delta\omega+\I\Gamma/2}{(\Delta\omega)^2+\Gamma^2/4}\,\mu f_{T}\,.
\end{equation}
We can now insert this solution into Eq.~(\ref{eq:hBmotion}),
providing us with the steady-state production rate of $S$ quanta,
\begin{equation}\label{eq:gmotion5}
\dot h_S=\frac{\Gamma\,\mu^2}{(\omega_A-\omega_S)^2+\Gamma^2/4}\,\,\,
\frac{1}{e^{\omega_A/T}\pm1}\,.
\end{equation}
The only assumption has been that $\mu$ is small compared with
$\Gamma$. Therefore, this solution also applies when many
oscillations take place between collisions as long as the
oscillation amplitude is small. While oscillations show up in our
original transient solution when at $t=0$ the ensemble had been set
up in a pure flavor, they disappear in the steady-state solution
which describes the average of the entire ensemble, not individual
particles.

We now specialize to the production of longitudinal HPs from the
mixing with L-plasmons. Therefore, $\omega_A=\omega_{\rm P}$,
$\Gamma=\Gamma_{\rm L}$ (damping rate for on-shell L-plasmons) and
mixing energy $\mu=-\chi m/2$. For $\omega=\omega_S$ near
$\omega_{\rm P}$ this result is identical with
Eq.~(\ref{eq:Lproduction}). However, it is based on the flavor
evolution of on-shell L-plasmons and is not applicable for
$\omega$ very different from $\omega_{\rm P}$.

\section{Emission from the Sun}

\subsection{Resonant emission}

We are here mostly interested in low-mass HPs with $m\ll\omega_{\rm
P}$. Moreover, all over the solar interior we have $\Gamma_{\rm
L}(\omega_{\rm P})\ll\omega_{\rm P}$ and so the HP production is
narrowly concentrated around $\omega=\omega_{\rm P}$ and we may
approximate the emission rate as the delta-function of
Eq.~(\ref{eq:Lproduction}). Under this approximation the energy-loss
rate per unit volume is
\begin{equation}
\label{eq:Q1}
Q=\frac{\chi^2 m^2}{e^{\omega_{\rm P}/T}-1}\,
\frac{\omega_{\rm P}^3}{4\pi}\sim \chi^2 m^2\,\frac{T\,\omega_{\rm P}^2}{4\pi}
=\chi^2 m^2\,\frac{\alpha\,T\,n_e}{m_e}\,.
\end{equation}
The approximate expression derives from expanding the exponential
because L-plasmons are highly occupied (in the solar interior
$\omega_{\rm P}/T$ does not exceed 0.23), causing perhaps a 10\%
overall error.

Integrating the emission rate over the standard solar model AGSS09ph~\cite{Serenelli:2009yc} (without
expanding the exponential in the emission-rate formula), we find for
the HP luminosity in the L-channel
\begin{equation}
L_{\rm S_L}=5.7\times 10^{21} \chi^2 \left(\frac{m}{\rm eV}\right)^2\, L_\odot\,.
\end{equation}

\begin{figure}[t]
   \centering
   \includegraphics[width=0.6\columnwidth]{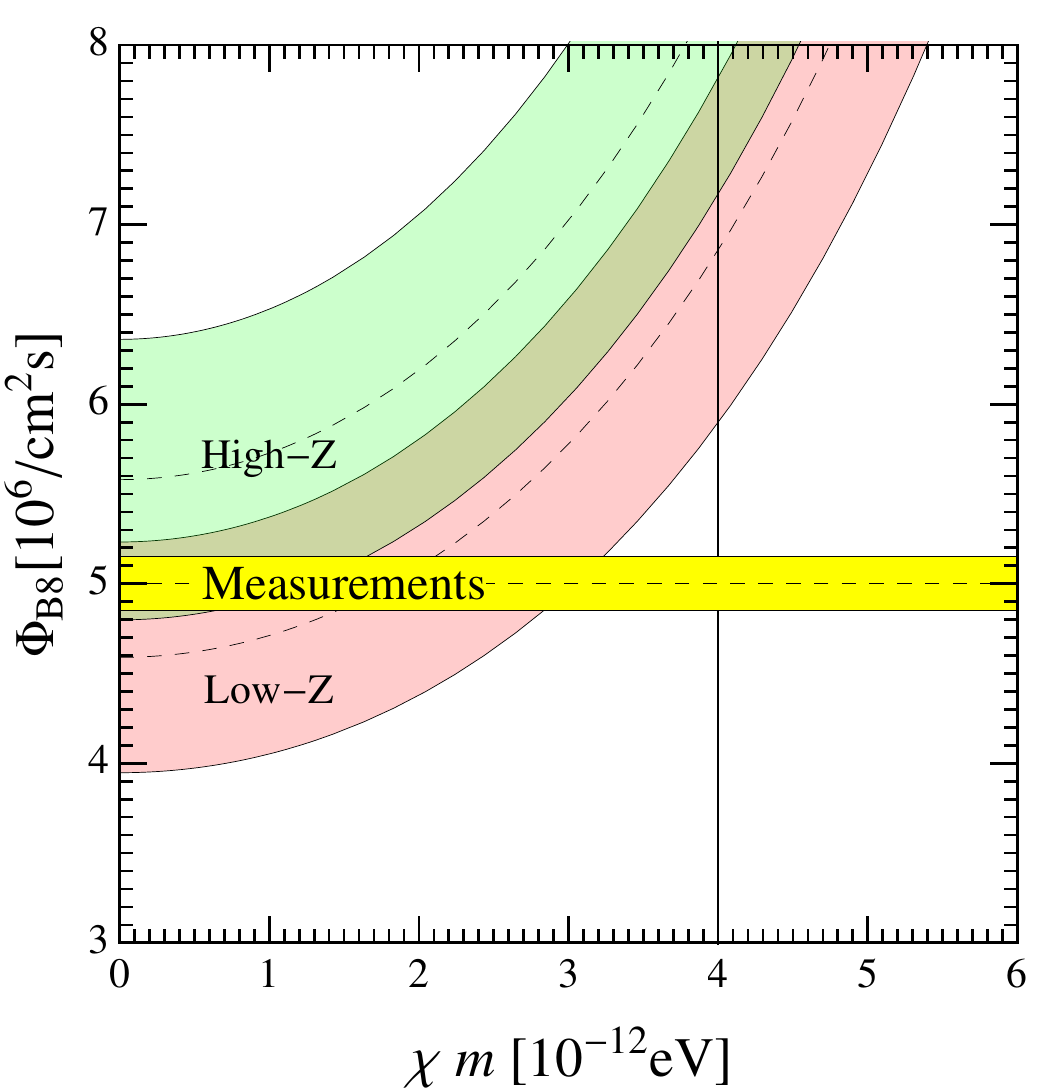}
   \caption{Solar $^8$B neutrino flux. {\it Yellow band}: Measurements. {\it Red band}:
   Expectation in the presence of longitudinal HP emission for a solar model with
   new opacities (low $Z$). {\it Green band:} Same for old opacities (high $Z$).
   The vertical line corresponds to $L_{S_L}<0.1\,L_\odot$ (our adopted limit).}
   \label{fig:Boron8}
\end{figure}

This exotic energy loss is constrained by our relatively precise
knowledge of the Sun. The most direct indicators of the properties
of the solar interior are the solar neutrino fluxes, and in
particular the boron neutrino flux which is especially sensitive to
temperature. It has been measured to be $\Phi^{\rm meas}_{\rm
B8}=5.00(1\pm 0.03)\times10^9/$cm$^2$s~ (see Table 2 of
Ref.~\cite{Robertson:2012ib}). The presence of some exotic energy
loss $L_x$ requires a larger nuclear generation rate and thus an
enhanced neutrino flux. It is given to good accuracy
by~\cite{Schlattl:1998fz, Gondolo:2008dd}
\begin{equation}
\Phi^x_{\rm B8}=\Phi^0_{\rm B0}\left(1+ L_x/L_\odot \right)^{4.6}\,,
\end{equation}
where $\Phi^0_{\rm B0}$ is the flux computed from a solar model
unperturbed by exotic losses. In Fig.~\ref{fig:Boron8} we show the
theoretical expectations for $\Phi^x_{\rm B0}$ for two solar models.

The low-$Z$ model (red band) is based on the latest studies of
the solar chemical composition~\cite{Asplund:2009fu}, which revealed
a lower metallicity (lower opacity and thus lower central solar
temperature) than previously assumed. It is in slight tension with
helioseismology, so for comparison we also show the the high-$Z$
model (green band), i.e., the earlier standard solar model, that
fits better helioseismological data. The widths of the bands
represent other uncertainties as taken from Table~2 of
Ref.~\cite{Robertson:2012ib}. This ``solar opacity problem'' for the
moment remains unresolved. Conceivably, the correct abundance of CNO
elements in the inner Sun can be determined by future solar neutrino
measurements. Ironically, the measured boron neutrino flux (yellow
band) lies exactly in the overlap region of the nominal error ranges
of the two cases.

Figure~\ref{fig:Boron8} suggests $\chi < 3\times 10^{-12}\,{\rm
eV}/{m}$ as a limit. Of course, the solar prediction is dominated by
systematic issues and an interpretation of the nominal uncertainties
in the form of meaningful confidence levels is not available.
Therefore, we follow Ref.~\cite{Gondolo:2008dd} and adopt the
requirement $L_{S_L}<0.1\,L_\odot$, providing our nominal limit
\begin{equation}\label{eq:mainlimit}
\chi< 4\times 10^{-12} \,\frac{\rm eV}{m}\,.
\end{equation}
It is shown as a vertical line in Fig.~\ref{fig:Boron8}. By present
evidence, this is a conservative constraint. Conceivably it could be
improved in future if the solar opacity problem can be convincingly
settled.

\subsection{Off-resonance production}

Since the L-channel dominates HP emission in at least some range
of masses, one may wonder if laboratory experiments could be
sensitive to this flux, again caused by HP-photon mixing. These
might be more sensitive in the X-ray regime of some keV rather than
the sub-keV energies produced by on-shell L-plasmon conversion.
Therefore, we compute the expected L-HP spectral flux at Earth for
energies above the solar plasma frequency, i.e., for
$\omega\gtrsim 0.3$~keV. This will also help us evaluate the solar
constraint for HP masses above $0.3~$keV.

For this purpose we need the explicit damping rate for L-plasmons
which is dominated by inverse bremsstrahlung at low energies and
Thomson scattering at high energies. We find explicitly
\begin{equation}
\Gamma_{\rm L}=
\frac{64\pi^2\alpha^3\,n_e \sum_{Z} Z^2 n_Z}{3 \sqrt{2 \pi\,T}\,m_e^{3/2}\,\omega^3}
\,F\left(\frac{\omega}{T}\right)
+\frac{8\pi \alpha^2n_e}{3 m_e^2}\sqrt{1-\frac{\omega_{\rm P}^2}{\omega^2}}\,,
\end{equation}
where $n_Z$ is the density of nuclei of charge $Z$ (in the Sun
essentially protons and alpha particles) and
\begin{equation}
F(w) =(1-e^{-w})
\int_0^\infty\hspace{-7pt} dx\,x\,e^{-x^2} \int_{\sqrt{x^2+w}-x}^{\sqrt{x^2+w}+x}\frac{t^3 d t}{(t^2+y^2)^2}\,.
\end{equation}
Here, $y=k_{\rm s}/\sqrt{2 m_e T}$ with $k_{\rm s}$ a screening
scale, i.e., we model screening by representing the interaction
between electrons and nuclei as a Yukawa potential. This approach
neglects various other corrections, notably Sommerfeld enhancement,
Pauli blocking for partially degenerate electrons and
electron-electron bremsstrahlung, which is small in the
nonrelativistic limit because of the equal mass of the colliding
particles. We also ignore free-bound and bound-bound transitions.
Overall we estimate the strike on $\Gamma_{\rm L}$ to be at
most some~20\%.

In the energy range of interest, screening reduces the emission
rate at most by a few tens of percent and is partially compensated
by other neglected effects. If we ignore screening entirely
($y=0$), our result agrees with Ref.~\cite{An:2013yfc}. In this
limit we find analytically
\begin{equation}
F(w)=K_0(w/2)\,{\rm sinh}(w/2)\,,
\end{equation}
where $K_0$ is a modified Bessel function of the second kind.

\begin{figure}
   \centering
   \includegraphics[width=0.9\columnwidth]{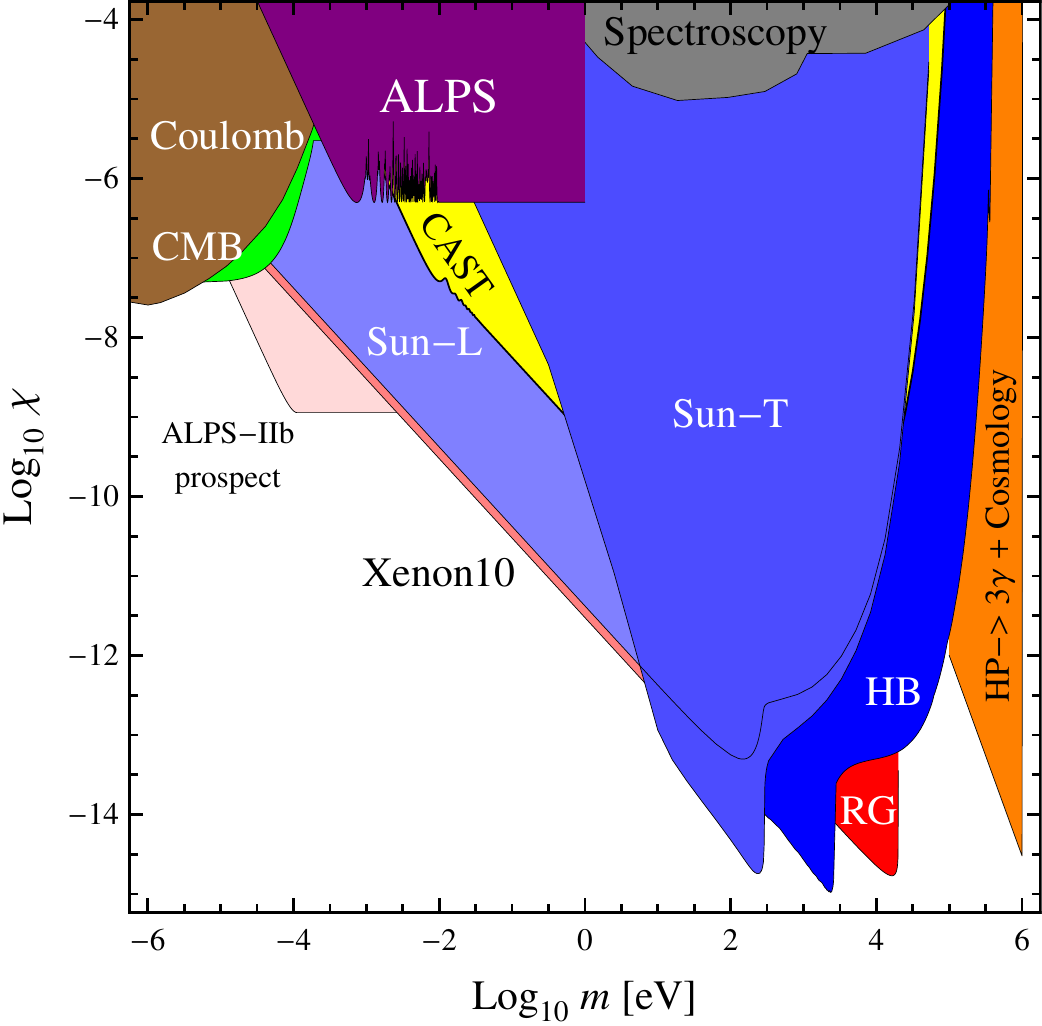}
   \caption{Bounds on hidden photons. The solar energy-loss constraints
   in the L and T channels as well as the bounds based on horizontal branch
   (HB) stars and red giants (RG) were derived here. We also show the CAST~\cite{Redondo:2008aa}
   and recent XENON10~\cite{An:2013yua} limits on solar HPs (see also~\cite{Mizumoto:2013jy, Horvat:2012yv}) as well as
   limits from modifications of Coulomb's law~\cite{Bartlett:1970js},
   distortions of the CMB spectrum~\cite{Jaeckel:2008fi}, the ALPS photon-regeneration experiment~\cite{Ehret:2010mh},
   atomic spectroscopy~\cite{Jaeckel:2010xx} and from decays of relic dark
   matter HPs~\cite{Pospelov:2008jk, Redondo:2008ec}. Also shown are prospects for the ALPS-II experiment~\cite{Bahre:2013ywa}.}
   \label{fig:bounds}
\end{figure}

It is now straightforward to express the energy-loss rate in terms
of $\Gamma_{\rm L}$ and integrate over the solar model
AGSS09ph~\cite{Serenelli:2009yc} for different HP masses. Imposing
the earlier constraint $L_{\rm HP}<0.1 L_\odot$ for the L-channel
leads to the constraints marked ``Sun-L'' in  Fig.~\ref{fig:bounds}.

For T-modes, exactly the same expressions pertain for $\Gamma_{\rm T}$
to lowest order in the electron velocity, in good agreement with
Ref.~\cite{Redondo:2008aa}.
Integrating over the same solar model and imposing the same
constraint leads to the limits marked ``Sun-T'' in
Fig.~\ref{fig:bounds}.

\subsection{Solar spectrum}

\begin{figure}[htbp]
   \centering
   \includegraphics[width=10cm]{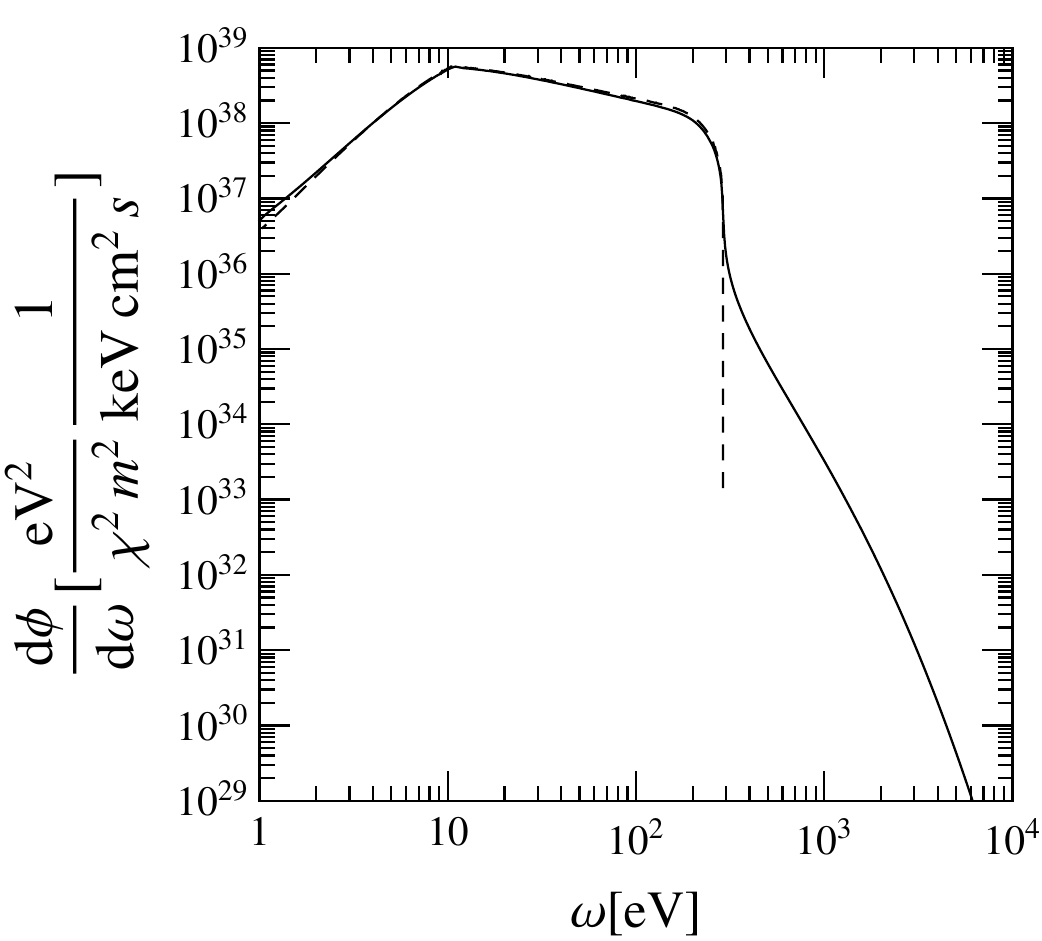}
   \caption{Flux at Earth of L-HPs in the limit $m\to 0$. For non-zero masses multiply
   with $\sqrt{1-m^2/\omega^2}$. The dashed line is
   the resonant flux based on the analytic arguments described in the text.}
   \label{fig:Lflux}
\end{figure}

For laboratory detection of solar HPs we need the spectral number
flux at Earth. The number of HPs emitted in the Sun per unit volume
and time is
\begin{equation}
\frac{d N}{dV dt }=\int\frac{d^3 {\bf k}}{(2\pi)^3}\Gamma_S^{\rm prod}\,,
\end{equation}
whereas the spectral flux at Earth is
\begin{equation}
\frac{d\Phi}{d\omega}=\frac{1}{(1~{\rm AU})^2}\int_0^{R_\odot} dr\,r^2
 \frac{\omega\sqrt{\omega^2-m^2}}{2\pi^2} \Gamma_S^{\rm prod}\,.
\end{equation}
We have performed this integral in the $m\to 0$ limit by using the
solar model AGSS09ph~\cite{Serenelli:2009yc} and obtain the flux
shown in Fig.~\ref{fig:Lflux}. We have included a screening
correction to bremsstrahlung based on the Debye scale including both
electrons and ions, causing a barely visible modification on the
scale of this plot.

The emission is dominated by energies below the maximum solar
$\omega_{\rm P}\sim 0.3$~keV where the production is resonant. The
energy dependence of this resonant flux can be explored by using
Eq.~\eqref{eq:deltaapprox} to perform the $r$ integration. Thus
using $\Gamma_S^{\rm prod}\propto \delta[\omega-\omega_{\rm P}(r)]$
yields
\begin{equation}
\frac{d\Phi}{d\omega}\approx
\frac{\chi^2m^2}{(1~{\rm AU})^2}\frac{r_\omega^2 T_\omega}{2\pi^2}
\left| \frac{d \log \omega^2_{\rm P}(r)}{d r}\right|^{-1}_{r=r(\omega)}\,,
\end{equation}
where $r_\omega$ and $T_\omega$ are the values of the radius and
temperature where $\omega_{\rm P}=\omega$. We have plotted this
contribution as a dashed line in Fig.~\ref{fig:Lflux}. Low energies
correspond to emission regions near the solar surface whereas
$\omega\to 0.3$~keV corresponds to the solar center. The analytical
formula describes well the numerical results, including the break at
$\omega\sim 10$ eV, which originates from the electron density
dropping much faster near the surface.

Energies $\omega\gtrsim 0.3$~keV cannot be produced resonantly and
the flux reduces considerably. An approximation formula for the flux
at Earth in this range is
\begin{eqnarray}
\frac{d\Phi}{d\omega}&=&\chi^2 m^2~5.7\times 10^{33}~{\rm cm}^{-2}~{\rm s}^{-1}~{\rm keV}^{-1}
\nonumber\\
&&{}\times\left[1+\frac{0.002}{(\omega-0.28)^{1.8}}\right]\,\omega^{-4}\,e^{-\omega/1.7}\,,
\end{eqnarray}
where $m$ is in eV and $\omega$ in keV. The accuracy is better than
10\% for $0.3~{\rm keV}<\omega<11~{\rm keV}$. The strong suppression
at high energies comes from the suppression of the mixing by medium
effects and the $\omega^{-3}$ dependence of the bremsstrahlung rate.
Above a few keV, the spectrum is exponentially suppressed.

\subsection{Direct detection of solar HPs}

The solar HP flux may be detectable in laboratory experiments. In
the T-channel, the CAST experiment provides the most restrictive
limit for sub-eV mass HPs (yellow region in Fig.~\ref{fig:bounds}).
However, the T-channel flux is maximum at eV~energies where 
the most sensitive CAST detectors are blind. 
Experiments aiming at $\sim$ eV HPs have already been
performed by CAST~\cite{Cantatore:2010zzb}, SUMICO~\cite{Mizumoto:2013jy} or 
are taking data
(SHIPS~\cite{Schwarz:2011gu}) but  their constraining power is
limited until a solid estimate of the low energy T-channel flux is
available (see also~\cite{Gninenko:2008pz,Redondo:2012ky}).

In the L-channel, large-scale dark matter detectors can be sensitive
by virtue of the inverse processes that produce HPs in the
Sun~\cite{An:2013yua}. At present, only the XENON10 experiment
appears to have any meaningful sensitivity.  According to
Ref.~\cite{An:2013yua}, the absence of excess counts above
background implies a limit $\chi<3\times 10^{-12}\,{\rm eV}/{m}$,
identical with the solar energy-loss limit suggested by our
Fig.~\ref{fig:Boron8} and slightly more restrictive than our adopted
conservative limit. In other words, dark matter detectors are only
beginning to probe solar HPs, leaving open the possibility that
future larger-scale instruments could actually find solar~HPs.

\section{Other stars}

To compare the impact of L-HP emission on stars other than the Sun,
it is easier to compare the usual energy-loss rate per unit mass. We
first focus on small HP masses for which some analytical insight can
be provided. The mass density is $\rho$ and the electron density is
approximately $n_e=Y_e \rho/m_p$, where $Y_e$ is the number of
electrons per baryon (1 for hydrogen, 0.5 for helium, carbon, and
oxygen). For small $m$, Eq.~(\ref{eq:Q1}) leads to
\begin{equation}
\varepsilon=\frac{Q}{\rho}\sim\chi^2 m^2\,\frac{\alpha}{m_e m_p}\,Y_e\,T\,.
\end{equation}
Essentially $\varepsilon$ depends only on temperature $T$.

Besides the Sun, one may consider HP emission from the
non-degenerate helium-burning cores of horizontal-branch (HB) stars
in globular clusters that usually provide more restrictive
constraints than the Sun, for example on axion emission. In the Sun,
a typical $T$ is 1~keV, in the cores of HB stars 8~keV. (Helium
burns at higher temperature than hydrogen.) $Y_e=0.5$ in HB-star
cores, in the Sun more like $0.8$. In the Sun, the average nuclear
energy generation rate is around $2~{\rm erg}~{\rm g}^{-1}~{\rm
s}^{-1}$, whereas in the helium-burning HB core it is around
$80~{\rm erg}~{\rm g}^{-1}~{\rm s}^{-1}$. In both cases, energy loss
into a ``dark channel'' is constrained to be less than some 10\% of
$\varepsilon_{\rm nuc}$. In other words, the constraint on HP
emission from the Sun would be roughly a factor of 10 more
restrictive. This unusual result arises from $\epsilon_{\rm HP}$ not
depending on density at all and on temperature only linearly.

However, for HP masses exceeding a few keV, the emission from the
Sun is thermally suppressed, providing a window where HB stars still
provide a useful limit. We have computed the HP luminosity (in L and
T modes) of a typical HB star (see Fig.~2.4
of~Ref.~\cite{Raffelt:1996}, taken from Ref.~\cite{Dearborn:1989he})
for different HP masses. Imposing the HP luminosity to be smaller
than 10\%, we obtain the constraint ``HB'' shown in
Fig.~\ref{fig:bounds}.

For yet larger HP masses, the higher density of more evolved stars
allows for resonant production. Red-giant stars before He ignition
probably provide an opportunity because in their degenerate core the
plasma frequency reaches $\omega_{\rm P}\sim 20$~keV at $T\sim
8.6$~keV. Imposing that the energy-loss rate in L-HPs is smaller
than $10$~erg~g$^{-1}$~s$^{-1}$ as suggested in~\cite{Raffelt:1996,
Raffelt:1989xu, Raffelt:1994ry, Catelan:1995ba} we obtain from
Eq.~(\ref{eq:Q1}) the exclusion region ``RG'' depicted in red in
Fig.~\ref{fig:bounds}.

For HP masses larger than 20~keV, we would need well studied stars
denser than RG cores before Helium ignition to produce HPs
resonantly, but no standard argument seems to be available.

\section{Production in the Early Universe}

In the low-mass region, the production of L HPs in the Sun
overshadows the production of T modes. It is reasonable to ask
whether this is also the case for the relic HPs produced in the big
bang, where L-modes were neglected in previous
works~\cite{Pospelov:2008jk, Jaeckel:2008fi, Redondo:2008ec}, at
least partially due to the mentioned error in the solar L-emission
rate~\cite{Redondo:2008aa}. An, Pospelov and Pradler have already
argued that L-modes are not copiously produced in the early
universe~\cite{An:2013yfc} and we agree with their arguments. Still,
it is instructive to provide a simple quantitative estimate. Since
HPs with $m>2m_e$ decay very efficiently into three photons, they
require extremely small $\chi$-values to be cosmologically stable
and we prefer to focus on $m<2m_e$.

The production of relic L-HPs follows neatly from our derivations
above. In particular, Eq.~\eqref{eq:Lproduction} holds because the
Hubble expansion is slow and thermal equilibrium is always a
good approximation. We define the usual comoving momentum as
$k_c=k a$, with $a=a(t)$ the scale factor. The phase-space
distribution of relic L HPs is given by integrating in time the
production rate as
\begin{equation}
f (k_c)= \int_0^{t_0}\Gamma_S^{\rm prod}(\omega,k_c/a)\,dt\,,
\end{equation}
where $\omega=\sqrt{(k_c/a)^2+m^2}$.

When $T\gtrsim m_e$, electrons and positrons are relativistic and
$\Pi_A$ has to be modified. However, even in this case it has
similar properties: Re$\,\Pi=m_\gamma^2\simeq \omega_{\rm P}^2$ and
Im$\,\Pi\ll$  Re $\Pi$. The exact form is not relevant. The
production of HPs with a given present-day momentum $k_c$ is
dominated by the resonance, i.e., when $\omega=m_\gamma\simeq
\omega_{\rm P}$ and the ambient on-shell L-plasmons can
oscillate into L-HPs. This simplifies the calculation, providing
\begin{equation}
f_L (k_c)\approx \left. \chi^2 \frac{\pi}{j(T)}\frac{ m^2 T}{\omega^2 H}\right|_{\rm res}
\end{equation}
where we have used $\exp (\omega/T)\sim \omega/T$. The Hubble factor
is $H=a^{-1}d a/dt$ and $j(T) = d \log \omega_{\rm P}^2/d \log a\sim
O(1)$ \cite{Redondo:2008ec}. On the RHS, $\omega$, $H$ and $T$ have
to be evaluated at the resonance.

The frequency dependence is $\sim T/\omega^2 H$, which strongly
decreases with $\omega$ (earlier resonances). Momenta with $k<m$
convert when $\omega_{\rm P}=\omega\simeq m$ and go through the
resonance almost simultaneously. These HPs are thus created
non-relativistically and constitute a form of dark matter. (They
decouple after the resonance if their mixing is small such that
indeed only the resonance produces them efficiently.)

The analogous relic density of transverse HPs
is~\cite{Redondo:2008ec}
\begin{equation}
f_T (k_c)\approx \left. \chi^2 \frac{\pi}{j(T)}\frac{m^2}{H T} \left[\frac{T}{\omega}\frac{1}{e^{\omega/T}-1}\right]\right|_{m=\omega_{\rm P}} .
\end{equation}
They feature an almost thermal spectrum and are produced resonantly
when $\omega_{\rm P}=m$, i.e., almost simultaneously with L-modes.
The abundance involves the evaluation of $H,T,j$ at the same moment
as L-modes so we can express our results as a ratio of densities,
\begin{equation}
\frac{n_{\rm L}}{n_{\rm T}}\simeq \frac{1}{\pi^2} \frac{m}{T_{\rm res}}\,.
\end{equation}
This ratio is always small because $m=\omega_{\rm P}$ is
suppressed with respect to $T$ at least by the electron charge. We
conclude that the dark matter or dark radiation in L-modes is always
smaller than in T-modes. As anticipated, the earlier DM estimates
based on T-modes alone~\cite{Redondo:2008ec} do not change
significantly for $m<2m_e$.

\section{Conclusions}

We have revisited solar emission of hidden photons. A previous
calculation by one of us~\cite{Redondo:2008aa} had missed a crucial
wave-function renormalization factor as correctly pointed out by An,
Pospelov and Pradler \cite{An:2013yfc}. We have derived the emission
rate in terms of the imaginary part of the in-medium HP self-energy
in the spirit of thermal field theory, similar to the approach
chosen in Ref.~\cite{An:2013yfc}, and using a kinetic approach,
closer to the picture of active-sterile flavor oscillations with
collisions, similar to the original approach of
Ref.~\cite{Redondo:2008aa}. In the resonant case, where ambient
on-shell L-plasmons convert to L-HPs, both results are identical.
For HP energies $\omega$ exceeding the plasma frequency, resonant
production is not possible and one needs the thermal field theory
approach.

We have updated several stellar energy-loss and cosmological limits,
but the only tangible change occurs for $m <  3$~eV where the most
restrictive limit among all astrophysical arguments arises from
solar L-mode emission. The measured $^8$B neutrino flux tightly
constrains the solar interior temperature and therefore the allowed
range of invisible energy losses $L_x$. Based on a generous upper
limit of $L_x<0.1\,L_\odot$ we have derived a new limit on the
kinetic mixing parameter given in Eq.~(\ref{eq:mainlimit}) and shown
in Fig.~\ref{fig:bounds} marked ``Sun-L.''

In future, large-scale dark-matter detectors may be able to find
solar HPs which cause ionization by bound-free transitions in the
detector material. At present, only XENON10 is marginally sensitive
to the solar flux and provides a constraint similar to the one
derived from the solar neutrino flux~\cite{An:2013yua}. Therefore,
future large-scale dark matter detectors have the opportunity to
detect solar HPs---any increase in sensitivity explores uncharted
territory in parameter space. Likewise, the photon regeneration
experiment ALPS-II, a pure laboratory approach, will explore a
region of low-mass HP parameter range that is apparently not
accessible by any other method.

\section{Acknowledgements}
We thank Alexander Kartavtsev for informative discussions about
thermal field theory. We acknowledge partial support by the
Deutsche Forschungsgemeinschaft through grant No. EXC 153 and by the
European Union through the Initial Training Network ``Invisibles,''
grant No.\ PITN-GA-2011-28944. J.R.\ acknowledges support by the
Alexander von Humboldt Foundation.


\end{document}